\documentclass[prb,
superscriptaddress,showpacs,amsmath,amssymb]{revtex4}

\begin{document}

\title{Black hole thermodynamics and topology}

\author{G.E.~Volovik}
\affiliation{Landau Institute for Theoretical Physics, acad. Semyonov av., 1a, 142432,
Chernogolovka, Russia}

\date{\today}

\begin{abstract}
Recently the difference between the Gibbons-Hawking temperature $T_{\rm GH}$ attributed to the Hawking radiation from the de Sitter cosmological horizon and the twice as high local temperature of the de Sitter state, $T=H/\pi=2T_{\rm GH}$, has been discussed by Hughes and Kusmartsev from the topological point of view (see arXiv:2505.05814). According to their approach, this difference is determined by the Euler characteristic $\chi({\cal M})$ of the considered spacetime with Euclidean time. The invariant $\chi({\cal M})$ is different for the global spacetime ${\cal M}=S^4$ and for the manifold limited to a region near the horizon, ${\cal M}=D^2\times S^2$. Here we consider the application of the topological approach to Reissner-Nordstr\"om (RN) black holes with two horizons. Both the outer and inner horizons are characterized by their near-horizon topology, which determines the corresponding horizon temperatures. As a result of the correlation between the horizons, the entropy of the RN black hole is independent of its electric charge, being completely determined by the mass of the black hole. This demonstrates the applicability of the topological approach to the multi-horizon systems.
\end{abstract}
\pacs{
}

\maketitle

\tableofcontents

\section{Introduction}

There is a difference between the temperature attributed to the Hawking radiation from the cosmological horizon in the de Sitter state \cite{GibbonsHawking1977} (the Gibbons-Hawking temperature $T_{\rm GH}=H/2\pi$) and the twice as high local temperature of the de Sitter state $T=H/\pi=2T_{\rm GH}$, see \cite{Volovik2025c,Maxfield2022} and references therein. There were different explanations of the factor 2.

Recently it was suggested that this difference has also a geometrical origin. In this topological approach, the temperature is determined by the Euler characteristic $\chi({\cal M})$ of the spacetime with the Euclidean time.\cite{Kusmartsev2025}
The topological Euler characteristic, which is determined by Chern–Gauss–Bonnet invariant, is different for different manifolds chosen for the consideration of the de Sitter thermodynamics. If the global de Sitter manifold $S^4$ is chosen, then its Euler characteristic  is  $\chi(S^4)=1$. If the manifold is considered, which is restricted to the near-horizon region outside the horizon, its topological invariant is
$\chi(D^2\times S^2)=2$. This gives the geometric support to the factor 2 difference between the two temperature. The temperature of the Hawking radiation from the cosmological horizon is determined by the near-horizon topology. This leads to the Hawking temperature, $\beta=\pi\chi(D^2\times S^2)/H= 2\pi/H$.
On the other hand, local processes, such as the ionization of an atom in the de Sitter environment, occur in bulk and have no relation to the horizon. They are described by the global $S^4$ topology, and the corresponding temperature of the de Sitter state is twice larger, $\beta=\pi\chi(S^4)/H= \pi/H$.

This geometric approach was also applied to the black holes,\cite{Liberati1997,Altas2023,Kusmartsev2025,Fairoos2025} where the topology of the near-horizon region outside the  black hole determines the temperature of the Hawking radiation. This approach was applied, in particular, to the Reissner-Nordstr\"om (RN) black hole with two horizons.\cite{Liberati1997,Fairoos2025} However, the obtained results were in contradiction with our results for the thermodynamics of RN black hole that the entropy of the RN black hole depends only on its mass.\cite{Volovik2021,Volovik2022,Singha2023}  Here we assume that this topological approach works, and reconsider its application to the RN black hole. The source of the contradiction is that only the topology associated with the outer horizon was considered, while the topology of the inner horizon was ignored. Here we include into consideration both the inner and outer horizons and obtain the values of the temperature and entropy of RN black hole in agreement with Refs.\cite{Volovik2021,Volovik2022,Singha2023}. This demonstrates that the topological approach can be applied to the thermodynamics of the RN black holes.

\section{Topology and temperature of inner and outer horizons}

We shall use expression for the Chern–Gauss–Bonnet (CGS) invariant $\chi({\cal M})$ in Ref.\cite{Fairoos2025}, but apply it not only to the outer horizon at $r=r_+$, but also to the inner horizon at $r=r_-$. These two horizons have different inverse temperatures, $\beta_+$ and $\beta_-$, each of them is determined by the corresponding CGS invariant. The invariant for the outer horizon:\cite{Fairoos2025}
\begin{equation}
\chi({\cal M})= \frac{1}{32\pi^2} \int_0^{\beta_+} d\tau \int_{r_+}^\infty dr \int_\Sigma \sqrt{g}d^2x 
\left(  R_{\mu\nu\rho\sigma}  R^{\mu\nu\rho\sigma} - 4 R_{\mu\nu}R^{\mu\nu}\right)\,,
\label{InvariantPlus}
\end{equation}
provides the connection between the radius $r_+$ and the corresponding temperature $\beta_+$.
In the same manner the topological invariant for the inner horizon must provide the connection between its radius $r_-$  and the corresponding temperature $\beta_-$:
\begin{equation}
\chi({\cal M})= \frac{1}{32\pi^2} \int_0^{\beta_-} d\tau \int_{r_-}^\infty dr \int_\Sigma \sqrt{g}d^2x 
\left(  R_{\mu\nu\rho\sigma}  R^{\mu\nu\rho\sigma} - 4 R_{\mu\nu}R^{\mu\nu}\right)\,.
\label{InvariantMinus}
\end{equation}

 For the RN black hole one has:
 \begin{equation}
 R_{\mu\nu\rho\sigma}  R^{\mu\nu\rho\sigma} - 4 R_{\mu\nu}R^{\mu\nu} = 
 \frac{8}{r^8}(6M^2r^2 -12MQ^2r +5Q^4)  \,,
\label{Curvature}
\end{equation}
where $r_+ + r_-=2M$ and $r_+ r_-=Q^2$ ($G=\hbar=1$).

With $\chi({\cal M})= 2$ the equation (\ref{InvariantPlus}) gives the temperature which is determined by the gravity at the outer horizon:
 \begin{equation}
\beta_+ =4\pi \frac{r_+^2}{r_+ - r_-} = \frac{2\pi  r_+^2}{\sqrt{M^2-Q^2}}\,.
\label{betap}
\end{equation}
In the same manner the equation (\ref{InvariantMinus}) gives the temperature $\beta_-$, which is determined by the gravity at the inner horizon:
 \begin{equation}
\beta_-=4\pi \frac{r_-^2}{r_- - r_+} =- \frac{2\pi  r_-^2}{\sqrt{M^2-Q^2}}\,.
\label{betam}
\end{equation}
This negative temperature is in agreement with Ref. \cite{ZhaiLiu2010}. Thus, the topological consideration is applicable to both the outer and inner horizons.

\section{Thermodynamics of black holes with two horizons}

The thermodynamics of RN black hole with two horizons is determined by both horizons. Example is provided by the Hawking radiation, which depends on both horizons.  It represents the coherent sequence of processes of tunneling across the two horizons, each determined by gravity at the corresponding horizon.  Such process is the analogue of the so-called co-tunneling -- the tunneling from an initial to the virtual intermediate states and then to final state state. The total radiation rate of this coherent process is
\begin{equation}
w \propto e^{-\beta_+ E} e^{-\beta_- E}=e^{-(\beta_+ + \beta_-) E} \,.
\label{beta}
\end{equation}
Thus the temperature, which determines the radiation rate from the RN black hole by co-tunneling process, does not coincide with the conventional Hawking temperature related to the outer horizon:
\cite{ZhaiLiu2010,Volovik2021,Volovik2022,Singha2023}
 \begin{equation}
\beta =\beta_+ +\beta_-=4\pi (r_+ + r_-)=8\pi M\,.
\label{beta}
\end{equation}
This temperature does not depend on charge of the RN black hole and is fully determined by its mass $M$. 

Accordingly, the entropy of a black hole also does not depend on the charge:\cite{Volovik2021,Volovik2022} 
\begin{equation}
S_{\rm RN}=4\pi M^2\,.
\label{entropy}
\end{equation}
This result can be obtained in several different ways. In particular, it is supported by the adiabatic process: the RN black hole can be adiabatically transformed to the Schwarzschild black hole by varying the fine structure constant $\alpha$ at fixed mass $M$. In this process the entropy does not change, and thus it is the same for RN black hole and for the Schwarzschild black hole.\cite{Volovik2022} This can be also obtained from the non-extensive Tsallis-Cirto $\delta=2$ statistics,\cite{TsallisCirto2013,Tsallis2020} which is applicable to the processes of merging and splitting of black holes,\cite{Volovik2025a} see also \cite{Manoharan2025}.
The modified Tsallis-Cirto $\delta=2$ statistics, which includes the negative entropy of the inner horizon, gives the following composition rule for entropy of the RN black hole in terms of the entropies of the two horizons:\cite{Volovik2025d}
 \begin{equation}
S_{\rm RN}= \left( \sqrt{S(r_+)} + \sqrt{|S(r_-)|}\right)^2= \pi(r_+ +r_-)^2 =4\pi M^2\,.
\label{entropyRN}
\end{equation}
Here $S(r_\pm) =\pm \pi r_\pm^2$.

Equations (\ref{entropy}) and (\ref{entropyRN}) are valid up to an extreme point, where $r_-=r_+$ and thus the two horizons merge and annihilate. At this transition point, continuity is broken,\cite{Carroll2009} and the adiabatic and topological approaches are no longer applicable. What happens in a supercritical regime, where instead of horizons there is a naked singularity, and whether a naked singularity is possible are open questions.\cite{Henheik2025,Joshi2025,Joshi2024,Virbhadra2022,YenChinOng2025,Dafermos2024} In Refs. \cite{Altas2023,Fairoos2025} it was suggested that for the extremal RN black hole the Gauss–Bonnet invariant is zero. If so, this means that at the transition point the topological invariant changes abruptly, and the entropy jumps from $S=4\pi M^2$ to $S=0$. Such jumps are typical in topological materials and in topological quantum filed theories. They take place at quantum phase transitions between states with different integer topological invariants.\cite{Volovik2007,Volovik2018} An example is the jump in quantized Hall conductivity during a quantum phase transition, in which the topological Chern number changes.
\section{Application to white holes}

This approach can be also applied to the horizon of the white hole. Using the anti-symmetry between the black and white holes (i.e. a white hole is time-reversal of the black hole), we can consider the white hole as a mirror of the black hole at negative $r$ (or a quantum clone in 't Hooft formulation \cite{Hooft2022,Hooft2025}). Then the topological invariant for the outer horizon of the white hole is:
\begin{equation}
\chi({\cal M})= \frac{1}{32\pi^2} \int_0^{\beta_+} d\tau \int_{-r_+}^{-\infty} dr \int_\Sigma \sqrt{g}d^2x 
\left(  R_{\mu\nu\rho\sigma}  R^{\mu\nu\rho\sigma} - 4 R_{\mu\nu}R^{\mu\nu}\right)\,.
\label{InvariantWhite}
\end{equation}
This gives for the inverse temperature $\beta_+$ of a white hole, the equation (\ref{betap}) with a minus sign:
$\beta_+({\rm white})= - \beta_+({\rm black})<0$. Correspondingly for the inner horizon of a white hole we have $\beta_-({\rm white})= - \beta_-({\rm black})>0$. As a result, the entropy of the white hole is with minus sign the entropy of the black hole with the same mass, in agreement with the anti-symmetry.\cite{Volovik2025d} 

The anti-symmetry of entropies is also supported by the process of macroscopic quantum tunnelling from black hole to white hole. The calculations of the tunnelling exponent gives the following rate of this process: $w\sim \exp(-2S_{\rm BH})$, where $S_{\rm BH}$ is the entropy of black hole. \cite{Volovik2022} The same result was obtained for the tunnelling of the black hole to white hole in Jackiw-Teitelboim gravity.\cite{Stanford2022} On the other hand, this quantum tunneling represents the rare quantum fluctuation, and thus its rate is determined by the difference between the entropies of these two macroscopic objects,  $w\sim \exp(\Delta S)$, where $\Delta S =S_{\rm WH}- S_{\rm BH}$. This gives $S_{\rm WH}=- S_{\rm BH}$.

\section{Conclusion}

It is shown that the topological approach to black hole thermodynamics and de Sitter spacetime proposed by Hughes and Kusmartsev\cite{Kusmartsev2025} also works for Reissner-Nordstr\"om black holes with two horizons. Both the outer and inner horizons with their near-horizon topology contribute to the thermodynamics of the RN black hole. The correlation between the horizons makes the entropy of the RN black hole independent of its electric charge, since it is completely determined by the mass of the black hole. The confirmation of the geometric approach is especially important for de Sitter thermodynamics. It supports the topological explanation of the difference between the Gibbons-Hawking temperature $T_{\rm GB}=H/2\pi$, which is determined by the near-horizon topology, and the local de Sitter temperature $T=H/\pi$, which is determined by the global de Sitter topology.\cite{Kusmartsev2025}

\end{document}